\documentstyle[12pt]{article}
\def\limfunc#1{\mathop {\rm #1}}

\begin{document}

\author{M. Chaves\\
Escuela de F\'{\i}sica\\
Universidad de Costa Rica\\
San Jos\'e, Costa Rica
}
\title{Some mathematical considerations about\\
generalized Yang-Mills theories
}
\maketitle

\begin{abstract}
Generalized Yang-Mills theories are constructed, that can use fields other
than vector as gauge fields. Their geometric interpretation is studied. An
application to the Glashow-Weinberg-Salam model is briefly review, and some
related mathematical and physical considerations are made.
\end{abstract}

\section{Introduction}

Yang-Mills theories have been tremendously useful in high energy physics,
where they have served to successfully model the electroweak and strong
interactions. The fundamental idea of a Yang-Mills theory is that its
mathematical expression must be invariant under a local compact Lie group of
transformations. By local here it is meant that the group element varies
with the point in Minkowski space that is being considered. In a typical
physical theory the quantum fields often appear differentiated with respect
to space or time. This means that, when the field transforms under the local
Lie group, the differentiation operator is going to act on the transforming
group element as well as on the field itself, so there is no invariance of
this term under the Lie group. Invariance is reestablished substituting the
differentiation operators by covariant derivatives. A covariant derivative
is the sum of the differential operator and the Yang-Mills field. These
fields are required to transform by adding terms that precisely cancel the
extra terms brought in by the differentiation operators. Such a
transformation is called a gauge transformation and the theory is left
unmodified by it. Fields that perform this kind of service are generically
called gauge fields.

Since the differentiation of a scalar field $\partial _\mu \varphi $, $%
\partial _\mu \equiv \partial /\partial x_\mu $, $\mu =0,1,2,3$, results in
a four-vector, the gauge fields have also been taken to be four-vectors. In
this article we are going to review\cite{max-herb} how it is possible to
generalize the concept of a covariant derivative in a Yang-Mills theory, so
that fields other than vector can be used as gauge fields.\cite{others} The
idea of putting the leptons in a triplet and using the graded group $SU(2/1)$
goes back to Ne'eman and Fairlie.\cite{old} The Lie group $SU(3)$ almost has
the right quantum numbers to embed the Glashow-Weinberg-Salam (GWS) Model in
it, but not quite. The right group seems to be $U(3)$ in a special
representation. An extra boson appears, but it automatically uncouples from
the rest of the particles.\cite{max-herb} However, the emphasis of this
paper is on the mathematical aspect, rather than on high energy applications.

\section{The kinetic energy of a vector field transforming spinorially}

The quantum electrodynamics Lagrangian is 
\begin{equation}
{\cal L}_{QED}=\bar{\psi}iD\!\!\!\!/\psi -\frac 14F^{\mu \nu }F_{\mu \nu }
\label{QED}
\end{equation}
where $D_\mu \equiv \partial _\mu +ieA_\mu $ and $F_{\mu \nu }\equiv
\partial _\mu A_\nu -\partial _\nu A_\mu =-ie^{-1}[D_\mu ,D_\nu ]$, and
where we are using a metric $\eta _{\mu \nu }=\limfunc{diag}(1,-1,-1,-1).$
This Lagrangian is invariant under a local transformation group based on the 
$U(1)$ Lie group. If $U=e^{-i\alpha (x)}$ is an element of this group, then
the electrically charged fermion field transforms as $\psi \rightarrow U\psi
.$ The vector field is required to obey the gauge transformation law 
\begin{equation}
A_\mu \rightarrow A_\mu +e^{-1}(\partial _\mu \alpha )\,{,}  \label{vec_1}
\end{equation}
so that, recalling the definition of $D_\mu $, the covariant derivative must
transform as 
\begin{equation}
D_\mu \rightarrow UD_\mu U^{-1}\,{.}  \label{derivative}
\end{equation}

In the previous equation the derivative that is part of the covariant
derivative is acting indefinitely to the right. We call such operators {\em %
unrestrained,} while the ones that act only on the immediately following
object, such as the partial in (\ref{vec_1}), we call {\em restrained}, and
use a parenthesis to emphasize that the action of the differentiation
operator does not extend any further{\em .} We do not find it admissible to
have unrestrained operators in a Lagrangian, because, first, they are not
gauge invariant, and, second, what they could physically or mathematically
mean is not clear. While the operators $D_\mu $ and $D_\mu D_\nu $ are
unrestrained, the operator $[D_\mu ,D_\nu ]$ is restrained, and it is
entirely appropriate that the kinetic energy of a vector boson can be
constructed using this commutator. The way the commutator becomes restrained
is as follows: 
\begin{equation}
\lbrack D_\mu ,D_\nu ]f=\partial _\mu A_\nu f-A_\nu \partial _\mu f-\partial
_\nu A_\mu f+A_\mu \partial _\nu f=(\partial _\mu A_\nu )f-(\partial _\nu
A_\mu )f,  \label{Leibnitz}
\end{equation}
where $f=f(x)$ is some differentiable function and it is seen how four
unrestrained operators result in two restrained ones, thanks to Leibnitz'
rule.

Let $S$ be an element of the spinor representation of the Lorentz group, so
that, if $\psi $ is a spinor, then it transforms as $\psi \rightarrow S\psi
. $ Then, due to the homomorphism that exists between the vector and spinor
representations of the Lorentz group, we have that $A\!\!\!/\rightarrow
SA\!\!\!/S^{-1}$. It was this homomorphism that allowed Dirac to write a
spinorial equation that included the vector electromagnetic field. The first
step in our road to a generalization of the covariant derivative will be to
rewrite the Lagrangian (\ref{QED}) using the spinorial representation. To
this effect we have the following

{\bf Theorem. }{\em Let }$D_\mu =\partial _\mu +B_\mu ${\em , where }$B_\mu $%
{\em \ is some vector field. Then: } 
\begin{equation}
(\partial _\mu B_\nu -\partial _\nu B_\mu )(\partial ^\mu B^\nu -\partial
^\nu B^\mu )=\frac 18\limfunc{Tr}^2D\!\!\!\!/^{\,2}-\frac 12%
\limfunc{Tr}D\!\!\!\!/^{\,4},  \label{teo_1}
\end{equation}
{\em where the traces are to be taken over the Dirac matrices.}

{\em Proof.} Notice the partials on the left of this identity are
restrained, the ones on the right are not. To prove the theorem it is
convenient to use the following trick, which makes the algebra manageable,
in this and in more complicated cases to follow. Consider the differentiable
operator $O\equiv \partial ^2+2B\cdot \partial +B^2$. Notice that it does
not contain any contractions with Dirac matrices, so that $\limfunc{Tr}O=4O,$
$\limfunc{Tr}O\left( \partial \!\!\!/B\!\!\!\!/\right) =4O\left( \partial
\cdot B\right) $, etc. It is not difficult to see then that $%
D\!\!\!\!/=O+\left( \partial \!\!\!/B\!\!\!\!/\right) $, where the slashed
partial is acting only on the succeeding slashed field. The trick is to use
this form of $D\!\!\!\!/^{\,2}$ in the traces on the right of (\ref{teo_1}).
With it one obtains 
\begin{eqnarray}
\frac 18\limfunc{Tr}^2\left[ O+\left( \partial
\!\!\!/B\!\!\!\!/\right) \right] -\frac 12\limfunc{Tr}\left[ O+\left(
\partial \!\!\!/B\!\!\!\!/\right) \right] ^2 &=&2\left( \partial \cdot
B\right) ^2-\frac 12\limfunc{Tr}\left[ \left( \partial
\!\!\!/B\!\!\!\!/\right) \left( \partial \!\!\!/B\!\!\!\!/\right) \right] 
\nonumber \\
&=&(\partial _\mu B_\nu -\partial _\nu B_\mu )(\partial ^\mu B^\nu -\partial
^\nu B^\mu )  \label{proof}
\end{eqnarray}
as we wished to demonstrate. The motivation for the additional trace term is
the same as for taking the commutator of the covariant derivatives: to
ensure that the differential operators be restrained.$\Box$

With the aid of the Theorem we can rewrite the QED Lagrangian in the form 
\begin{equation}
{\cal L}_{QED}=\bar{\psi}iD\!\!\!\!/\psi +e^{-2}\left( \frac 1{32}\limfunc{Tr%
}^2D\!\!\!\!/^{\,2}-\frac 18\limfunc{Tr}D\!\!\!\!/^{\,4}\right) 
\,{,}  \label{QED'}
\end{equation}
whose Lorentz invariance can be easily proven using $D\!\!\!\!/\rightarrow
SD\!\!\!\!/S^{-1}$ and the cyclic properties of the trace. As an example of
the invariance, observe that $\limfunc{Tr}D\!\!\!\!/^{\,2}\rightarrow 
\limfunc{Tr}SD\!\!\!\!/S^{-1}SD\!\!\!\!/S^{-1}=\limfunc{Tr}D\!\!\!\!/^{\,2}$.

\section{A scalar field functioning as a gauge field}

We are going to construct an example of a theory that employs a scalar
instead of a vector boson to maintain gauge invariance. To keep things
simple we use $U(1)$ as our Lie group, as in the previous section, and so,
again, (\ref{derivative}) must hold. This time, however, we take the
covariant derivative to be in the spinorial representation, as in (\ref{QED'}%
), so it becomes possible to define it to be: 
\begin{equation}
D_\varphi =\partial \!\!\!/-e\gamma ^5\varphi \,{.}  \label{der_phi}
\end{equation}
We now require the gauge field to transform as $\gamma ^5\varphi \rightarrow
\gamma ^5\varphi -ie^{-1}\partial \!\!\!/\alpha $. These equations
immediately assure us that $D_\varphi \rightarrow UD_\varphi U^{-1},$ and so
Lagrangian 
\begin{equation}
{\cal L}_\varphi =\bar{\psi}iD_\varphi \psi +e^{-2}\left( \frac 1{32}%
\limfunc{Tr}^2D_\varphi {}^2-\frac 18\limfunc{Tr}D_\varphi
{}^4\right)  \label{lag_phi1}
\end{equation}
is gauge invariant. The trace terms in this Lagrangian can be simplified
algebraically and the Lagrangian written in the more traditional form 
\begin{equation}
{\cal L}_\varphi =\bar{\psi}i\partial \!\!\!/\psi -e\bar{\psi}i\gamma
^5\varphi \psi +\frac 12(\partial _\mu \varphi )(\partial ^\mu \varphi )
\label{lag_phi2}
\end{equation}
after a bit of algebra. This last calculation is similar to the one done
last section, but with the $\gamma ^5$ taking the place of the $\gamma ^\mu $%
's of that previous calculation.

\section{Non-abelian Yang-Mills theory with mixed gauge fields}

Consider a Lagrangian that transforms under a non-abelian local Lie group
that has $N$ generators. The fermion or matter sector of the non-abelian
Lagrangian has the form $\bar{\psi}iD\!\!\!\!/\psi $, where $D_\mu $ is a
covariant derivative chosen to maintain gauge invariance. This term is
invariant under the transformation $\psi \rightarrow U\psi $, where $U=U(x)$
is an element of the fundamental representation of the group. The covariant
derivative is $D_\mu =\partial _\mu +A_\mu $, where $A_\mu =igA_\mu ^a(x)T^a$
is an element of the Lie algebra and $g$ is a coupling constant. We are
assuming here that the set of matrices $\{T^a\}$ is a representation of the
groups generators. Gauge invariance of the matter term is assured if 
\begin{equation}
A_\mu \rightarrow UA_\mu U^{-1}-(\partial _\mu U)U^{-1}\,{,}
\label{vec_2}
\end{equation}
or, what is the same, 
\begin{equation}
A\!\!\!/\rightarrow UA\!\!\!/U^{-1}-(\partial \!\!\!/U)U^{-1}  \label{vec_3}
\end{equation}
We have already seen how scalar fields can function as gauge fields. Our aim
in this section is to construct a non-abelian theory that uses both scalar
and vector gauge fields. We proceed as follows. For every generator in the
Lie group we choose one gauge field, it does not matter whether vector or
scalar. As an example, suppose there are $N$ generators in the Lie group; we
choose the first $N_V$ to be associated with an equal number of vector gauge
fields and the last $N_S$ to be associated with an equal number of scalar
gauge fields. Naturally $N_V+N_S=N$. Now we construct a covariant derivative 
$D$ by taking each one of the generators and multiplying it by one of its
associated gauge fields and summing them together. The result is 
\begin{equation}
D\equiv \partial \!\!\!/+A\!\!\!/+\Phi  \label{der_na}
\end{equation}
where 
\[
\begin{array}{ll}
A\!\!\!/=\gamma ^\mu A_\mu =ig\gamma ^\mu A_\mu ^aT^a, & a=1,\ldots ,N_V, \\ 
\Phi =\gamma ^5\varphi =-g\gamma ^5\varphi ^bT^b, & b=N_V+1,\ldots ,N.
\end{array}
\]

Notice the difference between $A_\mu $ and $A_\mu ^a$, and between $\varphi $
and $\varphi ^b$. We take the gauge transformation for these fields to be 
\begin{equation}
A\!\!\!/+\Phi \rightarrow U(A\!\!\!/+\Phi )U^{-1}-(\partial \!\!\!/U)U^{-1},
\label{vec_4}
\end{equation}
from which one can conclude that $D\rightarrow UDU^{-1}.$ The following
Lagrangian is constructed based on the requirements that it should contain
only matter fields and covariant derivatives, and that it possess both
Lorentz and gauge invariance: 
\begin{equation}
{\cal L}_{NA}=\bar{\psi}iD\psi +\frac 1{2g^2}\widetilde{\limfunc{Tr}}\left(
\frac 18\limfunc{Tr}^2D^{\,2}-\frac 12\limfunc{Tr}D^{\,4}\right)
\label{lag_na}
\end{equation}
where the trace with the tilde is over the Lie group matrices and the one
without it is over matrices of the spinorial representation of the Lorentz
group. The additional factor of 1/2 that the traces of (\ref{lag_na}) have
with respect to (\ref{teo_1}) comes from normalization $\widetilde{\limfunc{%
Tr}}T^aT^b=\frac 12\delta _{ab}$, the usual one in non-abelian gauge
theories.

Although we have constructed this non-abelian Lagrangian based only on the
requirements just mentioned, its expansion into component fields results in
expressions that are traditional in Yang-Mills theories. The reader who
wishes to make the expansion herself can substitute (\ref{der_na}) in (\ref
{lag_na}), keeping in mind the derivatives are unrestrained, and aim first
for the intermediate result 
\begin{eqnarray}
\frac 1{16}\limfunc{Tr}^2D^{\,2}-\frac 14\limfunc{Tr}D^{\,4}
&=&\left( (\partial \cdot A)+A^2\right) ^2-\limfunc{Tr}\left( (\partial
\!\!\!/A\!\!\!/)+A\!\!\!/A\!\!\!/\right) ^2  \nonumber \\
&&-\frac 14\limfunc{Tr}\left( (\partial \!\!\!/\Phi )+\{A\!\!\!/,\Phi
\}\right) ^2,  \label{inter}
\end{eqnarray}
where the curly brackets denote an anticommutator. (We recommend to use here
the trick explained in section 2.) Notice in this expression that the
differentiation operators are restrained, and that the two different types
of gauge fields appear only in an anticommutator. The $\gamma ^5$ in the
scalar boson term of the generalized covariant derivative $D$ ensures both
that the partials become restrained and that these anticommutators become
commutators once the properties of the Clifford matrices are taken into
account. Substituting (\ref{der_na}) in (\ref{inter}) and in the matter term
of (\ref{lag_na}) we obtain the non-abelian Lagrangian in expanded form: 
\begin{eqnarray}
{\cal L}_{NA} &=&\bar{\psi}i(\partial \!\!\!/+A\!\!\!/)\psi -g\bar{\psi}%
i\gamma ^5\varphi ^bT^b\psi +\frac 1{2g^2}\widetilde{\limfunc{Tr}}\left(
\partial _\mu A_\nu -\partial _\nu A_\mu +[A_\mu ,A_\nu ]\right) ^2 
\nonumber \\
&&+\frac 1{g^2}\widetilde{\limfunc{Tr}}\left( \partial _\mu \varphi +[A_\mu
,\varphi ]\right) ^2.  \label{lag_na'}
\end{eqnarray}
The reader will recognize familiar structures: the first term on the right
looks like the usual matter term of a gauge theory, the second like a Yukawa
term, the third like the kinetic energy of vector bosons in a Yang-Mills
theory and the fourth like the gauge-invariant kinetic energy of scalar
bosons in the non-abelian adjoint representation. It is also interesting to
observe that, if in the last term we set the vector bosons equal to zero,
then this term simply becomes $\sum_{b,\mu }\frac 12\partial _\mu \varphi
^b\partial ^\mu \varphi ^b$, the kinetic energy of the scalar bosons. We
have constructed a generic non-abelian gauge theory with gauge fields that
can be either scalar or vector.

\section{Applying these ideas to the GWS Model}

These ideas were applied\cite{max-herb} to the GWS Model of high energy
physics. The Higgs fields of the GWS Model were used, along with the usual
vector bosons, to construct a generalized covariant derivative. The original
intention was to use $SU(3)$ as the gauge Lie group, because it generates
quantum numbers for the particles that are very close to the experimental
ones. Eventually it was noticed that the complete leptonic and bosonic
sectors of the GWS Model could be written in the form of Lagrangian (\ref
{lag_na}) using the group $U(3).$

The covariant derivative $D$ contains the gauge vector bosons of $U(1)\times
SU(2),$ the scalar Higgs bosons, and a new scalar boson. The triplet $\psi
=(\nu _L,e_L,e_R)$ contains the leptons and is transformed by the
fundamental representation of $U(3)$. All quantum numbers are correctly
predicted, and an extra scalar boson, but it automatically decouples from
the rest of the model and is thus unobservable except through its
gravitational effects. The representation of the group generators is not the
usual one, but instead, a special one where the generators are obtained as a
linear combination of the usual ones.

The quarks have not been included so far into the scheme. There is one term
of the pertinent sector of the GWS Model that is not predicted by this
generalized derivative model, and it is the potential of the Higgs field $%
V\left( \varphi \right) $ that could cause the spontaneous symmetry breaking.

\section{A generalized curvature}

In a Yang-Mills theory, be it abelian or not, the terms with physical
content must be gauge invariant and not contain unrestrained derivatives.
For example, if $D_\mu $ is the covariant derivative of a non-abelian
theory, the usual expression for the kinetic energy of the gauge vector
bosons is $\widetilde{\limfunc{Tr}}[D_\mu ,D_\nu ][D^\mu ,D^\nu ],$ which
satisfies both conditions. But even the expression $F_{\mu \nu }=[D_\mu
,D_\nu ]$ by itself does not have any unrestrained derivatives, while $%
\limfunc{Tr}F_{\mu \nu }$ has the additional property of being gauge
invariant. This quantity is the curvature in a principal vector bundle. The
question arises if similar results as those for $F_{\mu \nu }$ hold also for
a theory with a generalized covariant derivative. The answer to this
question is in the affirmative. We proceed now to define a quantity which we
shall call the generalized curvature $F$. Let $D$ be the generalized
covariant derivative, as given in (\ref{der_na}); then: 
\begin{equation}
F\equiv {\bf 1}\frac 14\limfunc{Tr}D^{\,2}-D^{\,2},  \label{curvature}
\end{equation}
or else, in terms of the derivative's constituents, 
\begin{equation}
F={\bf 1}\partial \cdot A-(\partial \!\!\!/A\!\!\!/)+{\bf 1}A\cdot
A-A\!\!\!/A\!\!\!/-\partial \!\!\!/\Phi -\{A\!\!\!/,\Phi \},
\label{curvature'}
\end{equation}
where ${\bf 1}$ is a $4\times 4$ unit matrix, so that $\limfunc{Tr}{\bf 1=}%
4. $ It can be seen that there are no unrestrained derivatives, and clearly $%
\limfunc{Tr}F$ is gauge invariant.

In the case of a Yang-Mills theory the vector boson kinetic energy can be
written exclusively in terms of the curvature $F_{\mu \nu },$ the commutator
of the covariant derivative with itself. In the generalized case we study
here, the kinetic energy can also be written in terms of the quantity $F$
defined above, a quantity, that, as previous examples of the curvature, is
quadratic in $D$ and restrained. The relation between these two quantities
is 
\begin{equation}
\frac 18\limfunc{Tr}^2D^{\,2}-\frac 12\limfunc{Tr}D^{\,4}=-\frac 12%
\limfunc{Tr}F^2,  \label{curvature2}
\end{equation}
that uses only traces for the Dirac matrices.

\section{The curvature in terms of the covariant derivative}

Let us review the curvature concept using Riemannian geometry as example. It
is well-known that geodesic deviation and parallel transport around an
infinitesimally small closed curve in a Riemann manifold are two aspects of
the same construction, and that trivial results do not occur in each case
only for curved manifolds.\cite{Weinberg} Consider thus a four dimensional
Riemann manifold with metric $g_{\mu \nu }$ and a vector $B^\mu .$ The
parallel transport of a vector $B^\beta $ around a closed curve $C$ is given
by 
\begin{equation}
\Delta B^\beta =\oint_C\Gamma ^\beta {}_{\nu \sigma }B^\sigma {}\frac{dx^\nu 
}{ds}ds,  \label{path}
\end{equation}
where $s$ is a parametrization of the curve and $\Gamma ^\beta {}_{\nu
\sigma }$ is the connection in a Riemann space, the Christoffel symbol. Let
now $C$ be a small parallelogram made up of two short vectors $x^\mu $ and $%
y^\mu ,$ so that its area tensor is $\Delta S^{\mu \nu }=\frac 12x^{(\mu
}y^{\nu )}.$ The integral can be performed taking the vector field to be
constant along the sides of the parallelogram, and expressing the values of
the Christoffel symbol and the vector field at the curve as the first two
terms of an expansion about, say, the center of the parallelogram. Thus, if
the values of those quantities at a point in one of the sides of the
parallelogram are $\tilde{\Gamma}^\alpha {}_{\beta \delta }$ and $\tilde{B}%
^\beta ,$ then, for one of the sides, $\tilde{\Gamma}^\alpha {}_{\beta
\delta }=\Gamma ^\alpha {}_{\beta \delta }+\Gamma ^\alpha {}_{\beta \delta
,\sigma }\frac 12x^\sigma $ and $B^\beta +(B^\beta {}_{,\sigma }+\Gamma
^\beta {}_{\sigma \tau }B^\tau )\frac 12x^\sigma ,$ where the quantities
without tilde take their values at the center of the parallelogram. The
contribution of this side is: 
\begin{equation}
y^\delta \tilde{\Gamma}^\alpha {}_{\beta \delta }\tilde{B}^\beta
|_1=y^\delta (\Gamma ^\alpha {}_{\beta \delta }+\Gamma ^\alpha {}_{\beta
\delta ,\sigma }x^\sigma /2)[B^\beta +(B^\beta {}_{,\sigma }+\Gamma ^\beta
{}_{\sigma \tau }B^\tau )d^\sigma /2]  \label{side1}
\end{equation}
The contribution of this and its opposite side is: 
\begin{equation}
y^\delta \tilde{\Gamma}^\alpha {}_{\beta \delta }\tilde{B}^\beta
|_{1+3}=y^\delta dx^\sigma [\Gamma ^\alpha {}_{\beta \delta }B^\beta
{}_{,\sigma }+\Gamma ^\alpha {}_{\beta \delta }\Gamma ^\beta {}_{\sigma \tau
}B^\tau +\Gamma ^\alpha {}_{\beta \delta ,\sigma }B^\beta ].  \label{side34}
\end{equation}
Summing over the four sides one obtains 
\begin{equation}
\Delta B^\alpha =-R^\alpha {}_{\beta \gamma \delta }B^\beta x^\gamma
y^\delta =-R^\alpha {}_{\beta \gamma \delta }B^\beta \Delta S^{\gamma \delta
},  \label{deficit}
\end{equation}
where $R^\alpha {}_{\beta \gamma \delta }=\Gamma ^\alpha {}_{\beta \gamma
,\delta }-\Gamma ^\alpha {}_{\beta \delta ,\gamma }+\Gamma ^\alpha
{}_{\delta \tau }\Gamma ^\tau {}_{\beta \gamma }-\Gamma ^\alpha {}_{\gamma
\tau }\Gamma ^\tau {}_{\beta \delta }$ is the Riemann tensor. The positive
contributions to the tensor come from two opposite sides of the
parallelogram, and the negative from the other two sides, resulting in
structure of commutators that can be seen to arise from a commutator of the
covariant derivative with itself.

In the case of a Yang-Mills theory, the vector boson kinetic energy can be
written exclusively in terms of the curvature $F_{\mu \nu },$ the commutator
of the covariant derivative with itself. The parallel transport of a
Yang-Mills field, in its role as a connection, around a closed path\cite
{Rosner} results in the curvature, the Yang-Mills connection squared. For
two short vectors $x^\mu $ and $y^\mu $ that form a parallelogram parallel
transport results, as can be shown similarly to the way it was done in last
paragraph for the Christoffel symbol, in 
\begin{equation}
\oint A_\mu dx^\mu =F_{\mu \nu }x^\mu y^\nu ,  \label{YMcurvature}
\end{equation}
which gives a motivation for the interpretation of $F_{\mu \nu }$ as a
curvature.

In the generalized case we have been studying in this paper the kinetic
energy can also be written in terms of a curvature, precisely the quantity $%
F $ already defined, which was quadratic in $D$ and restrained. The equality
does not need the taking of the Lie group trace, and one simply has 
\begin{equation}
\frac 18\limfunc{Tr}^2D^{\,2}-\frac 12\limfunc{Tr}D^{\,4}=-\frac 12%
\limfunc{Tr}F^2.  \label{curvature4}
\end{equation}
This equality can be shown to be true substituting $F$ as given by (\ref
{curvature}) on the right in the equation above. Thus the kinetic energy
goes as the square of the curvature in the generalized case, too.

\section{Studying the geometric background}

Let us study the possible geometric interpretations of the formalism. Let
then $x^\mu $ and $y^\nu $ be two small vectors that form a parallelogram,
as before. To calculate the parallel transport, the dot product between
vectors has to be done taking a trace over the Dirac matrices, but otherwise
the procedure is the same as before. Using the generalized covariant
derivative and the contracted forms $x\!\!\!/$ and $y\!\!\!/$ for the two
vectors, and taking the small parallelogram as a path to perform the
parallel transport integral, one obtains 
\begin{equation}
\frac 14\limfunc{Tr}DDx\!\!\!/y\!\!\!/=F_{\mu \nu }x^\mu y^\nu ,
\label{partial}
\end{equation}
where $F_{\mu \nu }$ is constructed as usual with the vector fields that
make up the covariant derivative. While this result is reasonable, it is
unsatisfactory in that the scalar fields are left as useless bystanders.

One possible attempt to use the formalism we have developed to its fullest,
is adding other terms to $x\!\!\!/$ and $y\!\!\!/.$ Let then $x^5$ and $y^5$
be two numbers and define 
\begin{equation}
d_x\equiv x\!\!\!/+x^5\gamma ^5\quad{\rm and}\quad d_y\equiv y\!\!\!/+y^5\gamma ^5.
\label{5vectors}
\end{equation}
It is possible now to generalize the parallel transport integral, and
substitute (\ref{5vectors}) in the trace expression of (\ref{curvature4}).
The result is: 
\begin{eqnarray}
\frac 14\limfunc{Tr}DDx\!\!\!/y\!\!\!/ &=&\left( \partial _{[\mu }A_{\nu
]}+[A_\mu ,A_\nu ]\right) x^\mu y^\nu +\left( \partial _\mu \varphi +[A_\mu
,\varphi ]\right) \left( y^5x^\mu -x^5y^\mu \right)  \nonumber \\
&&+\partial _5A_\mu (x^5y^\mu -x^\mu y^5).  \label{curvature5}
\end{eqnarray}
The geometric interpretation of this equation is straightforward: we are
dealing with a five dimensional manifold. The metric is Tr$d_xd_y=4(x\cdot
y+x^5y^5).$

The previous result is interesting from a mathematical point of view, but it
does not seem to be very enlightening when it comes to an understanding of
the geometric meaning of the formalism as applied to the GWS Model. The
reason is that, once we have promoted the linearly independent term $%
x^5\gamma ^5$ in (\ref{5vectors}) to represent a new dimension, the
covariant derivative (\ref{der_na}) has to include an extra term, a
derivative with respect to the new dimension, and become $D\equiv \partial
\!\!\!/+\partial _5\gamma ^5+A\!\!\!/+\Phi .$ This could, in principle, add
several new terms to the generalized derivative, and we would not have the
GWS Model anymore. As it is, all the new terms vanish except one; but one
term alone is bad enough, and we conclude that this five dimensional model
cannot represent the GWS Model. For the record we present curvature due to
the generalized derivative with the extra dimension: 
\begin{equation}
F=\partial \cdot A-(\partial \!\!\!/A\!\!\!/)+A\cdot
A-A\!\!\!/A\!\!\!/-(\partial \!\!\!/\Phi )-\{A\!\!\!/,\Phi \}-(\partial
\!\!\!/_5A\!\!\!/);
\end{equation}
the sole remaining new term is the last one.

\section{Final remarks}

We have reviewed the construction of a generalized Yang-Mills theory that
uses fields other than vector as gauge fields. We have shown a possible
geometric interpretation that requires an additional length parameter, but
it requires, to be consistent, an additional term in the covariant
derivative with respect to the new dimension. The results are interesting
from a mathematical point of view, but not very helpful if one is trying to
understand the geometry of the formalism as applied to the GWS Model, since
this dimension is not observed.

In this model some of the generators of the Lie group are vectorial, while
other are scalar. From a mathematical point of view it make no difference
which ones are which, but if one is aiming to reproduce the GWS Model, one
is has no free choice in this matter. This is interesting because it relates
the symmetries of the base manifold to the internal symmetry space of the
particles, the Lie group. One would expect that eventually an association of
this kind should follow from first principles.

\end{document}